\documentclass[conference]{IEEEtran}

\usepackage{cite}
\usepackage{moreverb}
\usepackage{graphicx}
\usepackage{subfigure}
\usepackage{setspace}
\usepackage{xcolor}
\usepackage{algorithm2e}

\long\def\comment#1{}
\comment{

\journalname{XXXXXX}
\setcounter{page}{01}

\setlength{\textheight}{9.5in}

  \copyrightnote{This is an open access article distributed under the terms of the \href{https://creativecommons.org/licenses/by/4.0/}{Creative Commons Attribution license}, which permits unlimited use, distribution and reproduction in any medium so long as the original work is properly cited.}
  
\received{XXXX}
  \accepted{XXXX}
  \published{XXXX}
}

\begin{document}


\title{Toward A Network-Assisted Approach for Effective Ransomware Detection}



\author{
\IEEEauthorblockN{Tianrou Xia${^a}$, Yuanyi Sun${^a}$, Sencun Zhu${^a}$, Zeeshan Rasheed${^b}$, Khurram Shafique${^b}$ }
\IEEEauthorblockA{\\
${^a}$Dept. of Computer Science and Engineering, Penn State University, University Park, PA 16802, USA\\
${^b}$Novateur Research Solutions, VA 20147, USA \\
Emails: {\{tzx17, yus160, sxz16\}@psu.edu}, \{zrasheed,kshafique\}@novateurresearch.com}
}


\maketitle

\vspace{-2cm}

\begin{abstract} Ransomware is one kind of malware using cryptography to prevent victims from normal use of their computers. As a result, victims lose the access to their files and desktops unless they pay the ransom to the attackers. By the end of 2019, ransomware attack had caused more than 10 billion dollars of financial loss to enterprises and individuals. In this work, we propose a Network-Assisted Approach (NAA), which contains local detection and network-level detection, to help user determine whether a machine has been infected by ransomware. To evaluate its performance, we built 100 containers in Docker to simulate network scenarios. A hybrid ransomware sample which is close to real-world ransomware is deployed on stimulative infected machines. The experiment results show that our network-level detection mechanisms are separately applicable to WAN and LAN scenarios for ransomware detection.
\end{abstract}


\section{Introduction}
Ransomware is a type of malware which blocks computer users’ access to their data or systems by encrypting important files in computers. Victims have to pay the requested ransom to get decryption keys from the attackers so that they can recover their data and systems. Sometimes the files cannot be recovered even if ransom is paid either because by accident the victim destroys the file  which contains decryption key or because the attacker breaks promise. Since ransomware attack is easy to implement and attackers can extort a large amount of money once it succeeds, a lot of ransomware have emerged in recent years and caused huge losses worldwide.

Here are some examples of ransomware attacks. Petya \cite{petya} is a family of ransomware first discovered in March 2016. It targeted Microsoft Windows-based systems and encrypted a hard drive’s file system table to prevent the system from booting. Victims had to pay the ransom in Bitcoin in order to regain access to the system. In June 2017, a derivative of Petya called NotPetya \cite{petya} launched a global attack on Microsoft Windows systems again via EternalBlue exploits and totally caused more than 10 billion dollars financial losses. In October 2017, a new ransomware attack named Bad Rabbit~\cite{badrabbit} was discovered in Russia and Ukraine, which follows a similar pattern to Petya. It encrypted the Windows user's file tables and then demanded a Bitcoin payment to decrypt them. Some researchers believed that Bad Rabbit had been distributed due to a bogus update to Adobe Flash software. At that time, a lot of agencies were affected by this ransomware including Interfax, Odessa International Airport, Kiev Metro and the Ministry of Infrastructure of Ukraine. In 2018 and 2019, ransomware still played an important role in malware family and exerted a significant impact on global computer users, especially Microsoft Windows operating system users. GrandGrab, Hermes2.1, Ryuk, Scarab, LockerGoga, etc. are all ransomware emerged during these two years targeting at Microsoft Windows since this system is the most common operating system used by enterprises and organizations that are potential blackmail objects for whom large ransoms are affordable.

As Linux operating system becomes increasingly popular in recent years and more businesses than ever are running on Linux now, Linux-oriented ransomware have sprung up to attack Linux users for exorbitant profits. In 2017, KillDisk \cite{killdisk} ransomware encrypted files, demanded bitcoin ransoms and left Linux systems unbootable. Erebus \cite{erebus} ransomware affected about 3400 of NAYANA’s clients via malware-containing advertisements. In 2019, Lilocked \cite{lilocked} ransomware targeted Linux servers and gain root access to encrypt the files with extensions such as PHP, HTML, CSS, etc. The victims were guided to dark web to make a payment in bitcoin in order to recover their files. The mechanism behind this ransomware is still a secret, researchers are looking out for a sample to discover the solution for decrypting affected files. Compared with ransomware targeted for Microsoft Windows operating system, Linux-oriented ransomware have not made a huge impact on enterprises and individuals up to now. However, this situation could change in the near future because the ransomware makers are always driven by profits. It is inevitable that more companies and individuals in industry will adopt Linux system due to its security, stability and open-source-ness, which will lead to the generation of many ransomware targeted at Linux operating system.

Among all types of ransomware in ransomware family, cryptoworm is one of the most troublesome genre. It spreads in the form of a worm, which means it can replicate itself and spread to other computers. Thus, cryptoworm can produce more serious consequence than other kinds of ransomware from the overall point of view once it is successfully designed and put into use by attackers. WannaCry \cite{wannacry} is an example of cryptoworm which broke out in May 2017. It used EternalBlue exploits to gain accesses to Microsoft Windows operating systems. As soon as the cryptoworm infected a computer, it encrypted data on the computer and later extorted Bitcoin cryptocurrency from the victim. Many organization systems were infected and helped spread WannaCry at that time because those systems did not apply newest patches released by Microsoft. This attack affected about 200,000 computers across 150 countries and resulted in total damages ranging from hundreds of millions to billions of dollars.

Since ransomware attacks emerge in endlessly, people all around the world is suffering from unanticipated threats to their property. To help individuals and collectives to get rid of this kind of financial loss, ransomware detection is an indispensable topic in study.

To mitigate the damage of ransomware attacks especially cryptoworm attacks, we proposed a Network-Assisted Approach (NAA) for ransomware detection, which combines local detection and network-level detection that successively give user a local report and a comprehensive report about respective detection result. The comprehensive detection report uses wisdom of the crowd to help computer users determine whether they are undergoing a ransomware attack so that they can take actions timely and avoid ransomware extortion.

We designed a local detection algorithm that is applicable on all kinds of operating systems and implemented a local detection mechanism prototype targeted at Linux system. In the local detection algorithm, we considered three features displayed on local hosts, among which there is a brand new feature never been used by previous works to the best of our knowledge, to generate a local report in an accurate and instant manner.

As for network-level detection, we adapted ant colony optimization algorithm to our problem and implemented an ACO-based Mechanism (ACOM) which sufficiently collects information from other machines so that a comprehensive report can be generated to help user determine whether the local host is attacked by ransomware or not. We also implemented a simple method named Broadcasting Mechanism (BM) which exhaustively collects information and used wisdom of the crowd to help user determine current safety state. These two network-level detection mechanisms are separately suitable for ransomware detection in WAN and LAN.

To estimate the performance of NAA, we established 100 containers in Docker and applied a Linux ransomware sample GonnaCry to simulative infected containers to mimic network scenarios. Then, we launched NAA in each container to achieve the evaluation of accuracy, message overheads and latency.

The main contributions of this paper are:
\begin{enumerate}
    \item[(1)] Propose a ransomware detection approach NAA especially targets at cryptoworm, which combines local detection and network-level detection to generate a report for user's reference.
    \item[(2)] Present a local detection algorithm applicable to all operating systems and implement a prototype on Linux system.
    \item[(3)] Apply ACO algorithm to network-level detection to implement a sufficient and reliable network-level detection mechanism ACOM to collect information from network.
    \item[(4)] Build a network scenario by establishing 100 containers in Docker and launching a ransomware sample on simulative infected machines to estimate the performance of NAA.
\end{enumerate}

The rest of this paper is organized as follows: Section 2 describes the background knowledge of both ransomware and ransomware detection approaches; Section 3 explains our motivations and generalizes the outline of NAA; Section 4 describes the design and implementation of our local detection mechanism; Section 5 describes the details of ACOM and BM; Section 6 evaluates NAA's performance from accuracy, message overheads and latency; Section 7 concludes the paper and discusses future work.
\section{Related Work}
\label{related}

Cryptographic ransomware is also called crypto ransomware, which always encrypts user files and then extorts users for cryptocurrencies before providing the decryption key. It is favored by attackers because digital currencies such as Ukash or Bitcoin and other cryptocurrency provide strong anonymity, making it difficult to trace and prosecute the perpetrators based on ransom payment transactions. 
Previous works on ransomware detection mainly focused on checking the features that are displayed due to ransomware behaviors on local hosts. And, they were designed for Microsoft Windows operating system because Windows is more vulnerable and is the target of most crypto ransomware. 

In 2015, Ahmadian et al. \cite{Ahmadian} proposed a comprehensive ransomware taxonomy and presented a connection monitor and connection breaker technique for detecting highly survivable ransomwares in the key exchange protocol step. 
In 2016, Paik et al. \cite{paik2016poster} proposed a storage-level detection method, which detects the existence of ransomware based on storage-access activities, e.g., number of files accessed and read/write frequency. Scaife et al. \cite{DBLP:conf/icdcs/ScaifeCTB16} presented an early-warning detection system that alerts users during suspicious file activity using a set of behavior indicators like entropy, file differences, magic bytes and read/write frequency. K. Cabaj et al. \cite{DBLP:journals/network/CabajM16} analyzed the behavior of a popular ransomware named CryptoWall and proposed two real-time mitigation methods using SDN-based algorithm. C. Moore \cite{DBLP:conf/ccc2/Moore16} investigated ransomware detection methods that implement canary files to monitor changes under folders and ascertained that canary files offer limited value as there is no way to influence the ransomware to access the folder containing monitored files. Sgandurra et al. \cite{DBLP:journals/corr/SgandurraMML16} presented a machine learning approach for dynamically analyzing and classifying ransomware. It monitors application behaviors and checks characteristic signs of ransomware including file extension, read/write frequency and function calls. 

In 2017, Y. Feng et al. \cite{feng2017poster} proposed a new approach based on deception and behavior monitoring to detect crypto ransomware with no loss. Their approach creates decoy files and makes ransomware operate on decoy files firstly, and then monitor the decoy files by checking whether they are encrypted by ransomware through the comparison of Shannon entropy, file type and sdhash between original files and changed files. Chadha et al. \cite{Chadha} discussed several machine learning algorithms for discovering DGA domains and analyzed their performance. Kirda et al. \cite{Kirda} presented a dynamic analysis system which automatically generates an artificial environment and detects when ransomware interacts with user data. In their system, entropy, removed files and read/write frequency are considered as monitor objects. Chen et al. \cite{DBLP:conf/racs/ChenKYK17} monitored the actual behaviors of software to generate API call flow graphs and used data mining techniques to build a detection model for decide whether the software is benign or is a ransomware. Kharraz et al. \cite{DBLP:conf/raid/KharrazK17} proposed a defense approach which maintains a transparent buffer for all storage I/O and then monitors the I/O request patterns of applications on a per-process basis for signs of ransomware-like behaviors including entropy, removed files, file extension and read/write frequency. 

In 2018, Khashif et al. \cite{8328219} presented a hybrid approach that combined static and dynamic analysis to generate a set of features that characterizes the ransomware behavior. This approach analyzes software binary code first, and then checks entropy, canary files, read/write frequency and function calls. Alaam et al. \cite{DBLP:journals/corr/abs-1802-03909} presented a detection tool which uses artificial neutral network and Fast Fourier Transformation (FFT) to develop a solution to ransomware detection by checking functions and frequency. Quinkert et al. \cite{DBLP:journals/corr/abs-1803-01598} presented a defense system that learns features of malicious domains by observing the domains involved in known ransomware attacks and then monitors newly registered domains to identify potentially malicious ones.  Moussaileb et al. \cite{DBLP:conf/IEEEares/MoussailebBPBCL18} presented a graph-based ransomware countermeasure which uses per-thread file system to highlight the malicious behaviors such as modification of canary files and accesses to large number of directories in a small time period. Morato et al. \cite{DBLP:journals/jnca/MoratoBMI18} proposed an algorithm that can detect ransomware action over shared documents by applying a network traffic inspection device between local users and shared volumes. The inspection device extracts SMB protocol commands through every access to the shared volumes it monitored and analyzes SMB traffic to determine whether the network volumes shared using SMB protocol is attacked by ransomware or not. 

In 2019, A. O. Almashhadani et al. \cite{8674751} demonstrated a comprehensive behavioral analysis of crypto ransomware network activities including DGA, SMB traffic and general traffic for detection of ransomware. Lee et al. \cite{8772046} proposed a method that utilizes an entropy technique to measure a characteristic of the encrypted files. Machine learning is applied for classifying infected-files-based file entropy analysis. 

The above literature covers almost all features that characterizes the ransomware behaviors. Our local detection mechanism also uses some of this kind of features to help determine whether a local host is a suspicious victim or not whereas a brand new feature "read/write pattern" is considered as well to help make accurate diagnosis on local hosts. Moreover, our approach contains network-level detection to offer more accurate detection results. While the papers mentioned above designed defense methods for Windows system, our prototype of the local detection mechanism targets at Linux system which is the next popular attack object of ransomware attacks although our approach is applicable on both Windows system and Linux system. We can easily derive a Windows version using the same design but different system libraries and tools.
\section{Our Network-Assisted Approach}

\subsection{Background Knowledge}

\subsubsection{Characteristics of Ransomware Behaviors}
Ransomware attacks always access victim’s operating system in some way and encrypt a large number of user files or system files or both automatically in a short time. During this process, the infected system performs differently from what it should be when there is no ransomware attack. This common trace provides various kinds of useful information that can be extracted from a suspected victim when a ransomware is running. Although there are some differences on key generation and key preservation strategies among different ransomware, we can still conclude the following common features that show so obvious distinctions between safe and infected circumstances that can be used for ransomware detection.

\begin{enumerate}
    \item [(1)] Keywords \\
    If a software is a ransomware, it probably contains some keywords that are commonly used in ransomware binaries. For example, “bitcoin”, “crypto”, “ransom”, etc. are common strings frequently appear in ransomware binaries. By inspecting software binaries, we can figure out some suspicious software even before ransomware attack happens.
    
    \item[(2)] Function Calls \\
    Since ransomware needs to encrypt files, it always calls functions related to cryptographic algorithms, including key generation, encryption and decryption functions. These functions may be written by the attacker or invoked from existing libraries. We can inspect binaries to locate the software that call these functions. 
    
    \item[(3)] Data Information \\
    Once a file is encrypted, we can observe some changes on this file. The file extension may be modified to a specific extension designated by the ransomware. The entropy of the file increases due to the randomness of data after encryption. The magic bytes of this file are different from original bytes because they are encrypted. Some files are even deleted since the ransomware created new files to store encrypted versions. All of these features provide useful information for ransomware detection.
    
    \item[(4)] Metadata Information \\
    Metadata information refers to some indirect information we can collect during ransomware attacks instead of information from file contents. Ransomware is an automatic program that encrypts a large number of files in a very short time in most cases due to super-fast computation speed of computers. So, when a ransomware is working, it accesses many files and directories, and then performs read and write operations on these files in short time periods, which leads to high file/directory access rate and high read/write frequency on a computer. This phenomenon also indicates a potential ransomware attack.
    
    \item[(5)] Network Traffic \\
    Some ransomware generate and store their keys on a remote server so that victims cannot figure out decryption keys without paying ransom. As for this kind of ransomware, it must contact the remote server to get encryption key during the attack. Thus, an unknown network traffic that is not produced by the user of the local host can be inspected, which helps ransomware detection.
\end{enumerate}

All these features listed above can be used to judge if the computer is in abnormal conditions and thus help determine whether there is ransomware working on this computer. However, one feature alone in consideration is insufficient for accurate detection results. So, most detection approaches pick several features and combine their checking results together to decide whether to alert user ransomware attack or not.

\subsubsection{Wisdom of the Crowd}
The wisdom of the crowd \cite{wisdom} is a collective opinion produced by a group of people instead of an individual. Some experiments showed that the collective knowledge of ordinary people is more precise than that of an expert. The reason for this phenomenon is that there is idiosyncratic noise associated with each individual judgment, and taking the average over a large number of responses will go some way toward canceling the effect of this noise. Thus, this notion has been applied to many social information sites such as Quora, Wikipedia and Yahoo! Answers. 

For high accuracy of ransomware detection results, we also use wisdom of the crowd in network-level detection of NAA to generate a comprehensive report for users to reference and to determine whether they need to do further actions to deal with potential ransomware attacks.

\subsection{Our Motivation}
As we can observe in ransomware attack cases, majority of ransomware have this property: Their appearance and diffusion are related to network. If one machine is infected by some ransomware, the others in the same local area network (LAN) are potential victims. That is because LAN is deployed by entities such as enterprises, laboratories and schools to interconnect computers within a small area. Computers in one LAN are often equipped with the same operating system and the same version. So, if a computer is attacked by ransomware via some exploit, the others are possibly attacked or going to be attacked because they share the same vulnerability. Worse, if the ransomware is a cryptoworm that actively scans the local network to compromise other machines, those computers in the same LAN are hence under high risk.
Even if in a wild area network (WAN), cryptoworm can spread in high speed because it is self-propagating, which means one infected computer can infect almost all computers communicated with it and result in fast increase of infected computers.

So, network related information that is corresponding to the conditions of other computers in network is very useful in ransomware detection especially in cryptoworm detection. However, existing approaches for ransomware detection only consider the characteristics of ransomware behaviors on local hosts as the parameters of their detection tools. One exception is the work\cite{DBLP:journals/jnca/MoratoBMI18} that analyzes file sharing traffic in a volume sharing scenario to detect possible ransomware. However, this work still did not have an eye on the security information of other computers in network. 

Motivated by the above observations, we propose a network-assisted approach which contains both local detection and network-level detection. Local detection is responsible for checking local features of a machine to make a preliminary diagnosis whereas network-level detection collects security conditions of other machines in a specific area to help determine whether the local host is infected or not. Since one machine can be in both LAN and WAN, our network-level detection has two separate schemes for these two scenarios. To achieve security conditions of other machines from WAN, we design an ACO-based Mechanism (ACOM), which uses ant colony optimization algorithm to efficiently collect maximum amount of information with minimum network resource consumption and report its detection result to the user. To obtain desired information from LAN, we design Broadcasting Mechanism (BM). It directly collects information of all machines in LAN and uses wisdom of the crowd to report a comprehensive detection result. With the information provided by ACOM and BM, NAA can accurately detect ransomware especially cryptoworm and help user judge whether the local host is infected or not.

\subsection{Outline of NAA}
NAA is a ransomware detection application that does both local detection and network-level detection. The local detection checks local features while the network-level detection collects security conditions of other machines from network to provide information for user to judgement whether the local host is attacked or not. Figure 1 shows the workflow of NAA. 

\begin{figure}[!htb]
	\centering
		\includegraphics[width=3.3in]{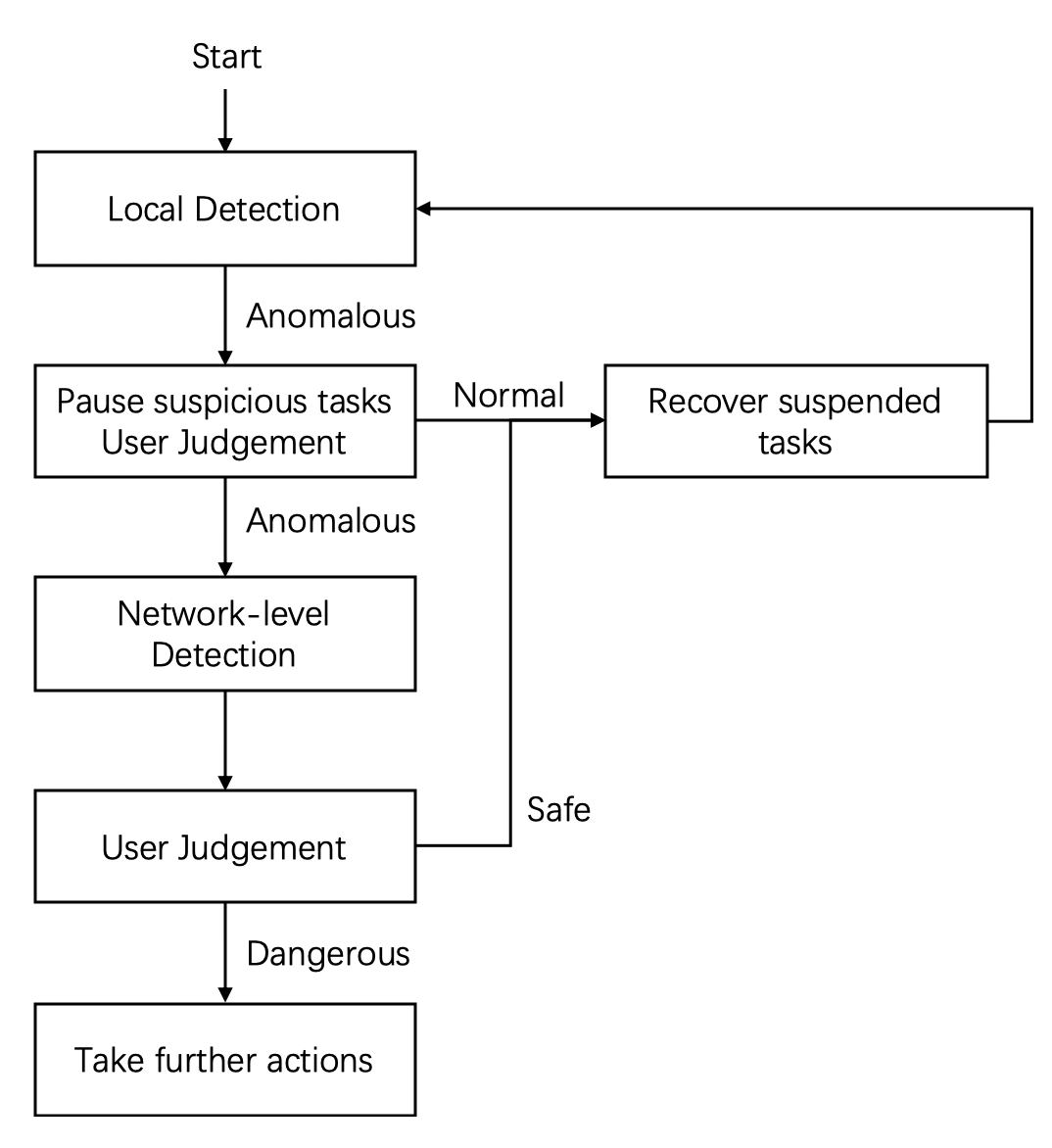}
	\caption{Workflow of NAA.}
\end{figure}

First of all, we run the local detection mechanism on each local host. If the local detection mechanism finds anomalous tasks, it suspends them using “kill -STOP pid” command and accordingly raises an alert to the end user. Then, based on his knowledge, the user should respond to NAA  whether the anomalous behaviors are caused by a legitimate user operation or not (e.g., when the user is encrypting files with a special tool). If they are, NAA will resume the suspended processes using "kill -CONT pid" command and continue to do local detection. If the user indicates these behaviors are anomalous (either because they are truly anomalous or because the user has no idea on what is going on), network-level detection should be launched. During the process of network-level detection, ACOM is responsible for collecting information from WAN and BM is responsible for collecting information from LAN. Once both mechanisms finish their work, a comprehensive report will be sent to the user describing the current network-wide situation. Then, the user can get an idea 
on the fraction of computers in the LAN that are also in the anomalous state and how ACOM views about the current state of this local host. Based on such given information, the user can make a judgement about whether this local host is in danger. 
If the answer is yes, NAA finishes its work; otherwise, the computer is considered safe and NAA will resume the suspended tasks and continue with local detection.


In the following sections, we will explain how local detection and network-level detection work and generate reports in detail.
\section{Local Detection}

\subsection{Design of Local Detection Algorithm}
Review that local detection checks some common features on local hosts which always display different characteristics under safe condition and infected condition. Section 3.1 introduced characteristics of ransomware behaviors and listed the features that could be considered in local detection.

In our local detection algorithm, we pick entropy, read/write frequency and read/write pattern as input parameters to diagnose the local host because the combination of these three parameters provides both high accuracy and efficiency. Among them, entropy and read/write frequency are classic features used by previous methods while read/write pattern is a brand new feature firstly proposed by this paper.

Entropy is the measurement of the randomness originally used in thermodynamics. In 1984, Claude E. Shannon applied entropy to digital communications in his paper “A Mathematical Theory of Communication” \cite{DBLP:journals/sigmobile/Shannon01}. After that, people started to use entropy to describe the extent of the randomness of a digital file. Encrypted files and compressed files tend to have higher entropy than normal files because the bytes in encrypted files and compressed files are more random. So, we can use entropy to help us determine if a file is in normal condition or not. However, even if we can find files with high entropy in a system, we cannot deem that this system is infected by ransomware because there are two exceptions: 1) The files with high entropy are compressed files instead of encrypted files. 2) The files are encrypted files, but they are encrypted by authorized users. Thus, this feature alone is not sufficient to produce an accurate ransomware detection result. We use further features to help us make more accurate judgements.

Read/write frequency describes the frequency of read and write operations on a machine. Ransomware always encrypts many files in a short time because they do not want to be detected before they finish work. Moreover, they want to encrypt as many files as possible so that the attackers are more likely to get ransom from the victim. We all know that file encryption task is related to read and write operations. So, if a ransomware is working, we can probably observe high read and write frequencies on a system. However, we still cannot make accurate diagnosis about whether a system is a potential victim or not with these two features because there are still some exceptions such as batch file compression. It has the same behaviors as ransomware attack when only considering entropy and read/write frequency.

To distinguish the behaviors of ransomware attack and other normal behaviors that also result in high file entropy and high read/write frequency such as batch file compression, we take read/write patterns as the third feature since different tasks usually have different read/write patterns. To our best knowledge, this feature has not been used in prior work. We use it to distinguish user's normal behaviors from ransomware activities. Here read/write patterns refer to the relationship between read and write operations occurred on a system. For example, if there is a read operation right after a write operation, we can use \{write, read\} to describe their relation during this period. If there is a read operation before write operation, but between them exists a close operation, we can use \{read, ..., write\} to describe the read/write pattern in this scenario which means there exist(s) other operation(s) between read and write. When ransomware is encrypting a file, it always reads the original file first and writes the ciphertext into a new created file. Then, the original file is deleted so that only encrypted file left. Ransomware encrypts files one after another, which makes read and write operations pairs appear at intervals. So, the pattern of ransomware activity can be concluded as \{read, write\}. In contrast, batch file compression task continuously reads each file in a specified directory and finally writes compressed texts into the compressed file after closing these original files. There is no adjacent read and write operations in its pattern. Other tasks also have their own read/write patterns that are usually different from those of ransomware activities. Thus, read/write patterns can help us filter out some benign behaviors when detecting ransomware attacks.

Our local detection algorithm comprehensively considers these three features to make a conclusion about whether the local host is anomalous or not. This algorithm is applicable to all operating systems because no matter what kind of operating system the ransomware is working on, it has common behaviors which will cause common characteristics. In our implementation, we used this algorithm to build a local detection mechanism prototype for Linux system as an example.

\subsection{Implementation of Local Detection Mechanism}
This subsection describes the outline and details of the local detection mechanism which is implemented to support network-level detection methods. Since the local detection algorithm described in Section 4.1 is applicable to any operating system, we selected Linux system as an example to implement a prototype.

\subsubsection{Overview}
As we already known, all ransomware encrypt files to extort victims. Thus, all ransomware activities are related to operations on file system. To monitor the related operations on Linux file system, we use a tool called inotify \cite{inotify} which is a Linux kernel subsystem that can monitor file system events and report changes. Inotify events include IN\_OPEN, IN\_ACCESS, IN\_MODIFY, IN\_DELETE and etc., among which IN\_ACCESS indicates read operation and IN\_MODIFY indicates write operation. We can use several system calls provided by the inotify API to monitor a specified directory. To monitor the entire file system, we can use "/." as the directory name to be monitored which represents the root directory of Linux file system. Once inotify starts to work, all events occurred in the directory tree can be captured and an event handler defined by us will deal with these events following detection requirements.

Our local detection mechanism prototype utilizes inotify to monitor Linux file system and combines altogether three features mentioned in Section 4.1 (entropy, read/write frequencies, read/write patterns) to measure whether the local host is anomalous or not. Figure 2 shows the workflow of the local detection mechanism.

\begin{figure}[!htb]
	\centering
		\includegraphics[width=3.3in]{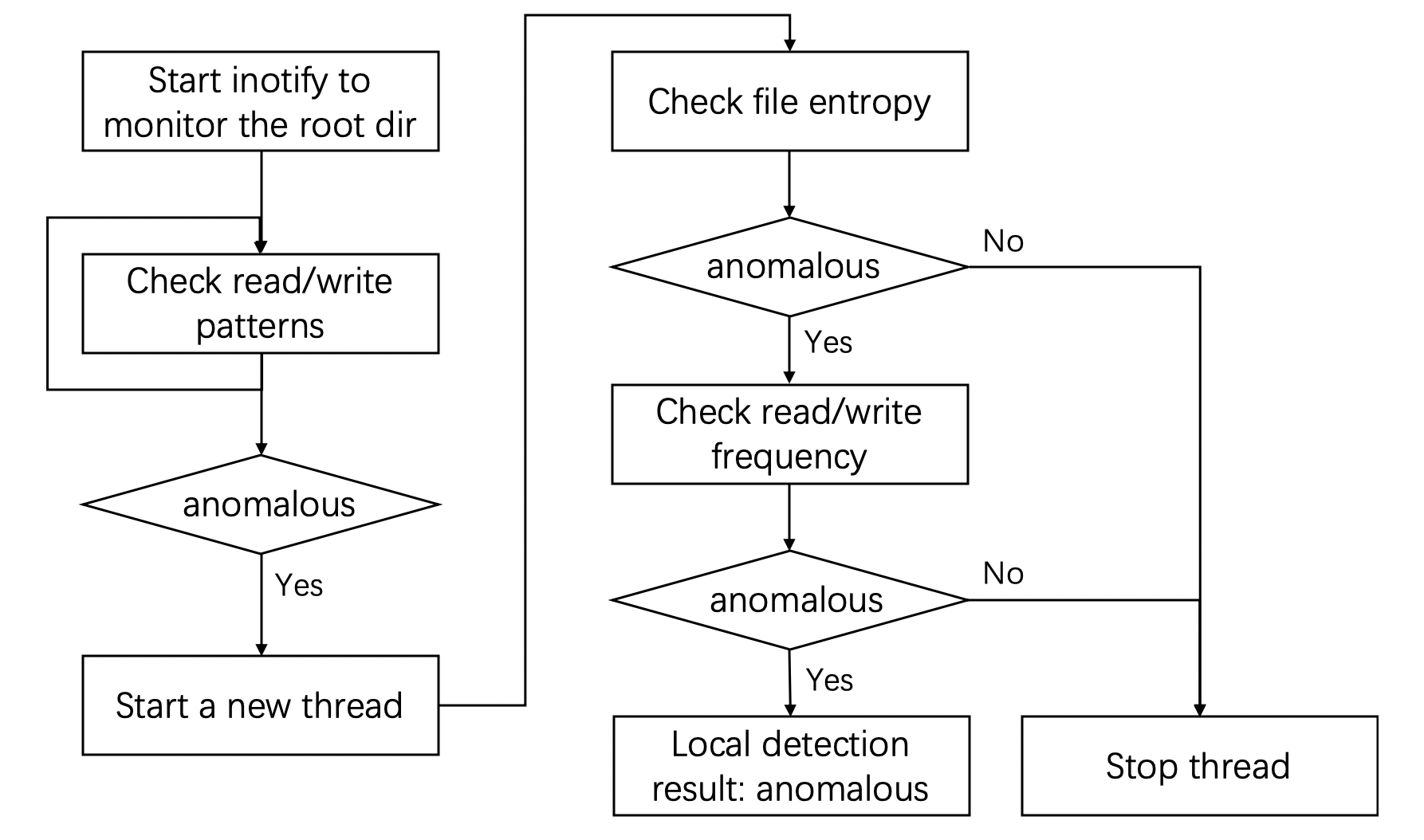}
	\caption{Workflow of local detection mechanism.}
\end{figure}

At the very beginning, we add a watch to the root directory so that we can monitor the entire file system. Then, start inotify. We first check read/write patterns because it can be done instantly when a new event is monitored. If there is a pattern matching with anomalous pattern, that is, the checking result of the first module is "anomalous", start a new thread to do further detection. This pattern checking module keeps working no matter what the result is because inotify keeps monitoring the file system and we don’t want to miss any possibly upcoming anomalous patterns. When we start the new thread, we also pass the path of the file where the anomalous write operation happened. 

The new thread works on checking the other features. It first checks file entropy of the potentially encrypted file whose path was passed by the pattern checking module when the new thread was created. If the file entropy is too high to be normal, that is, the checking result of the second module is "anomalous", go to the next module to check read/write frequency. Otherwise, the new thread stops because the local host is currently in safe state. Our reason for this judgement is that, the modified file, where the anomalous pattern is discovered, has normal entropy value which means it is not encrypted. This phenomenon is impossible to occur if the local host is undergoing a ransomware attack. In the third module, we check read/write frequency. If current read/write frequency on this system is too high to be normal, that is, the checking result of the third module is "anomalous", the local detection mechanism can make a diagnosis that this machine is anomalous because it shows anomalous characteristics in all three aspects. Otherwise, stop the new thread because the local host is safe. Note that, the local host has an initial state: safe. If the local detection mechanism cannot find the proof to confirm this machine is in anomalous state, we consider it is safe by default.

The rest of this subsection elaborates on how each module is implemented.

\subsubsection{Check Read/Write Patterns}
According to the work procedure of common ransomware, we know that ransomware always automatically encrypt files one after another. As for each file, the encryption task consists of several file operations: (1) Open the original file; (2) Create a new file; (3) Open the new file; (4) Read plaintext from the original file; (5) Write ciphertext in the new file; (6) Close the original file; (7) Close the new file; (8) Delete the original file. During this process, we can observe adjacent read and write operations with read before write. To distinguish read/write patterns of file encryption task with that of other tasks, we also observed the read/write patterns of some common user behaviors. By adding a watch to a particular directory, we can observe the events in this directory.

\begin{table*}[!htbp]
    \centering
    \small
    \begin{tabular}{ccc}
        \hline
        Tasks&File operations&Read/write patterns  \\
        \hline
        Encrypt a file&open,\ create,\ open,\ read,\ write,\ close,\ close,\ delete.&\{read, write\} \\
        Modify a file&open,\ read,\ close,\ open,\ create,\ open,\ close,\ write,\ close.&\{read, ..., write\} \\
        Compress a file&open,\ create,\ open,\ read,\ close,\ write,\ close.&\{read, ..., write\} \\
        Decompress a file&open,\ read,\ close.&\{read\} \\
        Browse a webpage 1&create,\ open,\ write,\ close,\ read,\ ...,\ read,\ write.&\{read*, write\} \\
        Browse a webpage 2&...,\ read,\ write,\ read,\ write,\ ...,\ read,\ write.&\{read, write\}* \\
        \hline
    \end{tabular}
    \caption{Read/write patterns of different tasks}
    \label{tab:table1}
\end{table*}

Table~\ref{tab:table1} lists file operations during file encryption and other normal tasks. According to this table, we can find that the read/write pattern of file encryption task is \{read, write\} which indicates a single pair of read and write operations with read before write. This \{read, write\} pair can appear many times, but other operations exist between two adjacent pairs. File modification and compression tasks have the following read/write pattern, \{read, ..., write\}, which means some other operations between read and write operations. When we decompress a file, only read operation occurs. The most confusing task is browsing a webpage, because it has similar read/write patterns as file encryption. When we browse a webpage, we can observe adjacent read and write operations as well. However, there exists continuous read operations before a write operation or iterative read/write pairs. So, we mark the read/write patterns of browsing a webpage as \{read*, write\} and \{read, write\}*, which are different from the read/write pattern of file encryption task.

Thus, we consider \{read, write\} as an anomalous read/write pattern indicating file encryption activities. Only when there is a read operation right before a write operation and before them are other file operations, we can say we find an anomalous pattern. We set a judgement condition that if there exists \{read, write\} on a monitored system, the local host is potentially in risk, further diagnosis is in need. Otherwise, the local host is safe. Since inotify monitors the entire file system in the implementation of our local detection mechanism, we admit that sometimes some operations from different tasks may mix together. That is, inotify may capture an operation from task A after an operation from task B but before another operation from task B, which may generate anomalous pattern while there is no anomalous behaviors. In this case, this pattern checking module causes false positives, that is why we need further diagnosis to check other features.

To be aware of the anomalous read/write patterns in time, we customize the inotify event handler in the following way: record all monitored events in order in an event\_list; once coming across a write operation, check the last two operations in event\_list. If the last one is read as well as the last-second one is neither read nor write, the anomalous read/write pattern is found; otherwise, empty the event\_list and continue to add monitored events into the list. Figure 3 shows the code of our event handler.

\comment{
\begin{figure}[!htb]
	\centering
	\footnotesize
	\begin{minted}{python}
# Handle inotify events
event_list = []
time_list = []
class MyHandler(pyinotify.ProcessEvent):
    def process_IN_CREATE(self, event):
        event_list.append("create")		 
    def process_IN_DELETE(self, event):
        event_list.append("delete") 
    def process_IN_MODIFY(self, event):		
        time_list.append(time.time())
        if event_list[-1] == "read" and \
           event_list[-2] != "read" and \
           event_list[-2] != "write":
            thread.start_new_thread(FurtherDiag,\
            (event.pathname,))
        event_list = []
        time_list = []
    def process_IN_OPEN(self, event):
        event_list.append("open")		
    def process_IN_ACCESS(self, event):
        time_list.append(time.time())
        event_list.append("read")
    def process_IN_CLOSE_WRITE(self, event):
        event_list.append("close")	
    def process_IN_CLOSE_NOWRITE(self, event):
        event_list.append("close")
	\end{minted}
	\caption{Event handler for local detection.}
\end{figure}
}

\begin{figure}[!htb]
	\centering
	\includegraphics[width=3.5in]{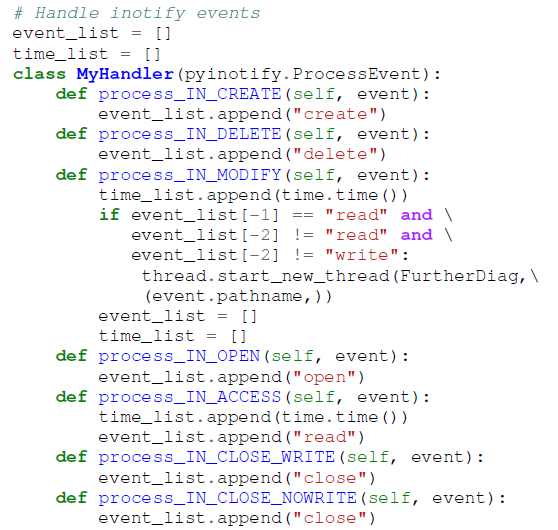}
	\caption{Event handler for local detection.}
\end{figure}

Once an anomalous read/write pattern \{read, write\} is discovered on a system, the checking result of the first module is "anomalous". So, we should start a new thread to do further diagnosis and pass the path of the file where this anomalous write operation happened to the new thread so that the second module can directly locate the file it needs to check.

\subsubsection{Check File Entropy}
There is an existing algorithm for file entropy calculation \cite{entropy}. Given a file, this algorithm traverses the target file to get the frequency count of each byte value and then uses the following formula to cumulatively calculate the entropy of the entire file.
\begin{equation}
    entropy = entropy + freq * \log_2 freq
\end{equation}
Here, the variable "entropy" is initialized to 0 and gradually increases until all "freq" related values are included, the variable "freq" represents the frequency of each byte value. With this algorithm, we can easily calculate final entropy value for a target file. 

To distinguish normal files and encrypted files through file entropy, we launched an experiment to calculate the entropy values of various kinds of normal files and encrypted files. Table~\ref{tab:table2} lists the entropy values of many different types of files in normal state and encrypted state.

\begin{table*}[!htbp]
    \centering
    \small
    \begin{tabular}{ccc}
        \hline
        File types&Normal state&Encrypted state  \\
        \hline
        .txt&4.62&7.98 \\
        .log&4.76&7.83 \\
        .conf&4.47&7.92 \\
        .pgn&7.91&8.00 \\
        .jpeg&7.94&8.00 \\
        .pptx&7.94&8.00 \\
        .mp3&7.95&8.00 \\
        \hline
    \end{tabular}
    \caption{File entropy of different files in normal state and encrypted state}
    \label{tab:table2}
\end{table*}

We can observe from Table~\ref{tab:table2} that text files which consist of English words have relatively low entropy in normal state. The entropy of this kind of normal files ranges from 4.0 to 5.0 while that of their corresponding encrypted files ranges from 7.0 to 8.0 in Linux file system. As for other types of files such like pictures and audios, they have relatively high entropy even in normal state. After being encrypted, their entropy values are tend to be 8. So, we deal with different kinds of files in different ways. As for a text file, we set the threshold 6.00. As for an non-text file, the threshold is set to be 7.99. Then, we can determine whether a file is anomalous or not by checking its entropy.

First, we check file extension of the target file. If the file extension is out of our knowledge, this file must be encrypted by ransomware because ransomware always modify file extension after encrypting a file. If we can recognize the file extension, calculate file entropy and compare entropy value with appropriate threshold value. If the entropy of the inspected file is greater than or equal to the threshold value, this file is considered to have an anomalous entropy value. That is, the checking result of the second module is "anomalous". Then, the third feature "read/write frequency" should be checked for final detection result. Otherwise, this is not an encrypted file, hence not a ransomware attack.

\subsubsection{Check Read/Write Frequency}
The final checkpoint concerns read/write frequency on the local host. Once a read or write operation is monitored by inotify, the event handler will record the time it occurred, as shown in Figure 3. What is more, the redundant contents in time\_list will be removed at the beginning of the new thread so that only the read and write operations that occurred after \{read, write\} pattern will be recorded in time\_list. Since we ran a new thread for further diagnosis, event handler can continue to record the time of upcoming read and write operations. With the recorded information in time\_list, we can calculate read/write frequency in the system after the anomalous pattern is found, which is defined as the average number of read/write operations occurred per second: 
\begin{equation}
    read/write\ frequency = \frac{operation\ counts}{duration},
\end{equation}
where "operation counts" represents the total number of recorded read and write operations after an anomalous read/write pattern, "duration" represents the time interval between the first recorded operation time and the last one in time\_list. We can achieve the value of "operation counts" by counting the number of elements in time\_list and calculate "duration" by computing the difference between the first and the last element in time\_list.

To distinguish normal read/write frequency with anomalous read/write frequency caused by ransomware activities, we did two experiments that respectively tests the read/write frequency during simulative ransomware activities and user normal behaviors. 

\begin{table*}[!htbp]
    \centering
    \small
    \begin{tabular}{cccccc}
        \hline
        \ &1\ KB&10\ KB&100\ KB&500\ KB&1\ MB  \\
        \hline
        AES\_128\_CBC&742\ op/sec&1379\ op/sec&8318\ op/sec&33876\ op/sec&43363\ op/sec \\
        AES\_256\_CBC&724\ op/sec&1437\ op/sec&8642\ op/sec&33920\ op/sec&43780\ op/sec \\
        AES\_128\_ECB&749\ op/sec&1440\ op/sec&8758\ op/sec&34162\ op/sec&43027\ op/sec \\
        AES\_256\_ECB&788\ op/sec&1380\ op/sec&8546\ op/sec&33697\ op/sec&43998\ op/sec \\
        RSA&651\ op/sec&-&-&-&- \\
        Op counts&200&400&2600&12400&24600 \\
        \hline
    \end{tabular}
    \caption{Read/write frequency during batch file encryption.}
    \label{tab:table3}
\end{table*}

\begin{table*}[!htbp]
    \centering
    \small
    \begin{tabular}{ccc}
        \hline
        Applications&Max Frequency&Average Frequency  \\
        \hline
        Firefox&322\ op/sec&95 op/sec \\
        Text\ editor&210\ op/sec&88 op/sec \\
        LibreOffice\ writer&310\ op/sec&35 op/sec \\
        YouTube&342\ op/sec&105 op/sec \\
        Amazon&281 op/sec&121 op/sec \\
        Gmail&253 op/sec&74 op/sec \\
        \hline
    \end{tabular}
    \caption{Read/write frequency during normal behaviors.}
    \label{tab:table4}
\end{table*}

In the first experiment, we use AES ciphers and RSA cipher from openssl library to encrypt files whose sizes range from 1KB to 1MB. As for each test, given cipher type and file size, encrypt 100 files automatically. Table~\ref{tab:table3} shows the experiment results. When the file size is specified, the read/write frequency hardly changes with different ciphers applied. When the cipher type is decided, larger files tend to cause larger read/write frequency. When we use RSA cipher, it can only encrypt small files due to the limitation of its encryption key length in openssl library, so, we did not get test results for relatively large files when RSA is applied. However, it does not matter because in real-world ransomware, RSA is always used to encrypt keys whose length is relatively small. In the tests, we also observed the number of read and write operations occurred during file encryption tasks. By analyzing the data in Table~\ref{tab:table3}, we found the read/write frequency on a system undergoing ransomware attack should be over 600 operations per second. Even if the ransomware is encrypting files smaller than 1 KB, the read/write frequency  could not be smaller than 600 op/sec. The reason is that, when the file size is 1 KB, there are totally 200 read and write operations happened on 100 files. That is to say, there is only one read and one write operation during the encryption of one file. So, when ransomware works on files that are smaller than 1 KB, the number of read/write operations will not change whereas the time consumption can be smaller than that of encrypting 1 KB files, which makes read/write frequency larger than 600 op/sec. Therefore, we can set the lower bound of the read/write frequency during ransomware activity to be larger than 600 op/sec.

Then, we use another experiment to test the read/write frequency during normal user behaviors. Table~\ref{tab:table4} shows the experiment results. For example, when we use Firefox, the maximum read/write frequency on this machine is 322 op/sec and the average read/write frequency is 95 op/sec. When we watch a video on YouTube, the maximum frequency is 342 op/sec while the average frequency is only 105 op/sec. We can observe that the upper bound of read/write frequency during normal user activities are smaller than 400 op/sec.

Since the upper bound of normal read/write frequency is lower than 400 op/sec meanwhile the lower bound of anomalous read/write frequency is higher than 600 op/sec. We picked the mid number 500 as the threshold. If the current observed read/write frequency is greater than or equal to 500 op/sec, the checking result of the third module will be "anomalous". Then, the local detection mechanism can finish its work with an "anomalous" detection result. Otherwise, since the read/write frequency is normal, this machine is considered safe. \\

In summary, the local detection mechanism uses inotify to keep monitoring the local host and checking read/write patterns. An anomalous read/write pattern will trigger further diagnosis. If all features show anomalous checking results, the local detection mechanism will send an alert to user reporting anomalous state on this machine and suspicious tasks that are performing anomalous behaviors. After that, all running tasks on this machine will be suspended and then the network-level detection will be triggered to collect information from other machines.

\subsection{Validation of Local Detection Mechanism}
As we mentioned in Section 3.1, using one feature alone to detect ransomware is not sufficient because single feature methods will cause many false positives and false negatives. For example, if we use file entropy as the only feature to determine whether a machine is infected, the compressed files will be mistaken for encrypted files and result in false positives. To validate the service of our local detection mechanism, we applied it on two machines under two different scenarios. 

In the first scenario, both of these two test machines are safe. We ran our local detection mechanism on them for two days and used them as usual such as doing course projects, reading papers, writing assignments, watching movies, playing computer games and etc. In the second scenario, we also ran our local detection mechanism on these two test machines for two days, but during this period, we applied the Linux ransomware GonnaCry \cite{GonnaCry} on them at random time for 48 times and recorded the detection results.

\begin{table*}[!htbp]
    \centering
    \small
    \begin{tabular}{ccc}
        \hline
        Machine&Number of false positives&Number of false negatives  \\
        \hline
        Machine1&3&0\\
        Machine2&0&0 \\
        \hline
    \end{tabular}
    \caption{False positives and false negatives caused by local detection mechanism}
    \label{tab:table5}
\end{table*}

Table~\ref{tab:table5} shows the test results, we can know that there were 3 false positives on Machine1 but no false negative case during the experiment. That is to say, when the test machines are in safe state, our local detection mechanism reported "anomalous" detection results for three times on Machine1. When the test machines are under the risk of ransomware attacks, all attacks were correctly detected and reported by our local detection mechanism. We also found the reason for these 3 false positives. They are caused by file encryption behaviors performed by authorized users. 

Sometime, although there is no ransomware attack, users' ransomware-like behaviors will cause false positives. That's why we need network-level detection to help us correct some false positives of local detection and to provide users with more accurate information to judge whether there is a ransomware attack indeed.
\section{Network-Level Detection}
The network-level detection works on collecting security conditions of other machines from network and generating a comprehensive report to help user determine whether there exists ransomware attack. It can help correct some false positives made by local detection and it enjoys excellent functionality especially when there is a cryptoworm attack.

The general idea of network-level detection is that, if multiple machines manifested the similar anomalous behavior at about the same time, it is likely a cryptoworm attack. If only a few machines are anomalous, these machines may be misdiagnosed by local detection because cryptoworm spreads swiftly, causing a mass of infected machines. It is easy to know the number of anomalous machines in LAN by collecting information from all the peers. However, this idea is hard to be put into practice in WAN because it is difficult to efficiently collect useful information. If we query all machines in WAN for their security conditions, it will be time and network-resource consuming. If we only pick several machines as representatives, their information may not be reliable because a few machines’ information cannot reveal the condition of the entire WAN. To solve this dilemma, our solution is to apply the ACO algorithm to the network-level detection so that we can more efficiently collect the most useful information in the least time. 

To sum up, we use ACO-based Mechanism (ACOM) to collect information from selected machines in WAN and use Broadcasting Mechanism (BM) to collect information from all machines in the same LAN. Then, we can use wisdom of the crowd to provide user with collected data for reference and help user determine whether to treat this machine as an infected one or not.

\subsection{ACO-based Mechanism}

\subsubsection{Ant Colony Optimization}
Ant colony optimization (ACO) is an optimization technique inspired by the path finding behaviors of ants searching for food \cite{ACO}. In nature, ants use pheromone to communicate with each other. They left pheromone along with the path they find food so that other ants can also find food following the pheromone trails. When there are multiple pheromone paths ahead, ants make decision depending on the strength of pheromone trails. Most ants choose the strongest pheromone trial and only a small number of ants choose other ways. Over time, pheromone trails will gradually evaporate. This means that pheromone trails which no longer lead to a food source will eventually stop being used, promoting ants to find new paths and new food sources. Figure 4 gives an example of how ants searching for food. 

\begin{figure}[!htb]
	\centering
		\includegraphics[width=3.0in]{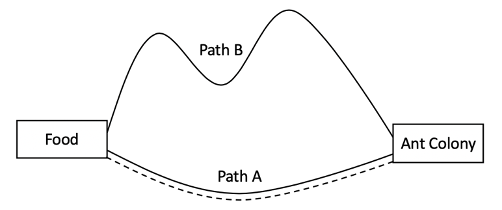}
	\caption{Path finding behavior of ants searching for food.}
\end{figure}

Suppose the food resource is on the left side and the ant colony is on the right side. There are two paths between food resource and ant colony. Path A has shorter distance while path B has longer distance. At the very beginning, both paths may be chosen by ants from the ant colony and pheromone trails are left on both paths. Since path A has shorter distance, the ants on path A spend less time to go and back which makes the pheromone trails on this path stronger than that on path B. The stronger pheromone trail on path A will attract more ants to this path. Overtime, almost all ants choose path A instead of path B. That is a process how ants find the shortest path between two places. So, ACO algorithm is always applied to optimization problems such as travelling salesman problem and various scheduling and routing problems. It has also been applied to detect network intrusions and Botnet servers~\cite{aco4botnet}.

Our problem is similar to travelling salesman problem. Instead of finding the shortest way to go through all cities, we want to find the shortest way to collect most information from other machines in WAN. So, we used ACO algorithm to help us do network-level detection in WAN scenario so that we can provide user with a helpful report without consuming too much network resources.

\subsubsection{Design of ACOM}
There are two key elements in ACO: ants and pheromone. To apply ACO to the network-level detection, we should first decide what roles these two elements should play in our approach. Since we want to collect most information from other machines in WAN, we use ants to collect and transmit information among machines just as what they do when searching for food. Each anomalous machine creates an ant and sends it to the network.  Each time an ant passes an anomalous machine, it records the security condition of this machine in it and share the information it has collected with the next machine it reaches. We consider pheromone as the number of anomalous machines each ant has collected, and it can be left on the machines that the ant passed. In this manner, as ants travel in WAN, machines can have increasing knowledge of the number of anomalous machines in WAN.

Then, according to the records in an ant when it finishes its work and the level of pheromone left on the machine, ACOM will generate a report telling user current situation in WAN. 
Figure 5 shows the pseudo code of ACOM, which describes the work procedure of this network-level detection mechanism.

\comment{
\begin{figure}[!htb]
	\centering
	\footnotesize
	\begin{minted}{python}
# start ACOM
CreateAnt() # Key function 1
Send ant to a randomly chosen machine
while True:
    Notify local host to do local detection
    ExchangeInformation() # Key function 2
    if num of anomalous machines known by ant >= goal:
        Ant goes back home
        Report: Alert.
        break
    else if num of anomalous machines known by ant < goal\
            and number of passed machines reaches limit:
        Ant goes back home
        Report: Low risk.
        break
    else:
        DecideDirection() # Key function 3
        Send ant to the selected machine
    \end{minted}
	\caption{Pseudo code describing the work procedure of ACOM.}
\end{figure}
}

\begin{figure}[!htb]
	\centering
	\includegraphics[width=4in]{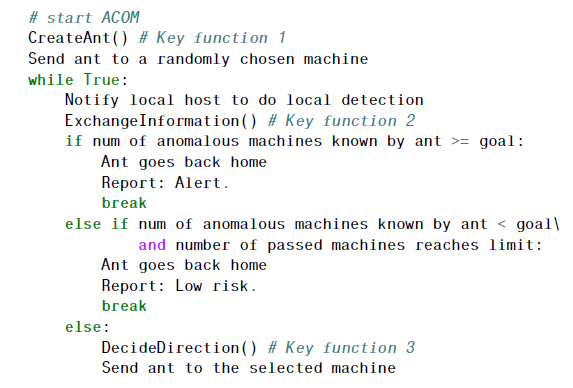}
	\caption{Pseudo code describing the work procedure of ACOM.}
\end{figure}

Once ACOM is launched, the anomalous local host creates an ant and then sends this ant to network. The next destination of the ant should be randomly selected from all machines this local host can contact with. Then, ACOM goes into a while loop. In this loop, the ant firstly notifies the current local host to do local detection again if this local host is not doing local detection. Then they exchange information with each other. The local host here indicates the machine that an ant is currently on. For example, we say machine A created an ant and sent it to machine B, the event “exchange information” happens between the ant and machine B. After information exchange, ACOM checks if the ant has achieved its goal which is the number of anomalous machines it needs to collect during its travel. If the ant has collected sufficient anomalous machines indicating a cryptoworm attack, it will go back to the original machine that created this ant and report to the user saying that "At least T users in WAN think you are in high risk". Here, T should be replaced by the value of threshold determined in different network environments. If the ant has not achieved goal but has reached the upper bound of its capability, it will go back as well but report that "We inquired 20 users in WAN, only A user(s) think(s) your are in risk." A should be replaced by the number of anomalous machines known by the ant. Both of the above two cases lead to the end of ACOM since it finished to provide user with wisdom of the crowd for reference. Otherwise, the ant should continue to work. The current local host it is on should decide the next stop of the ant according to pheromone information and send the ant to the next stop. The work procedure in the while loop iterates until the ant goes back to its original local host and reports our judgement. This is the entire workflow of ACOM. The detailed implementation of ACOM will be illustrated in the following subsection.

\subsubsection{Implementation of ACOM}
In the workflow of ACOM, there are three important functions: CreateAnt(), ExchangeInformation(), and DecideDirection(). The details of these three functions are explained below.

\noindent
\textbf{Key Function 1: CreateAnt()} \\
Ants are used to help the anomalous machines collect security condition information of other machines from network. In ACOM, anomalous machines create their own ants and send them to network to collect information of other machines. When a local host creates an ant, it needs to tell the ant three main things: goal, home, and (upper) limit. 

From a global perspective, we need to set a threshold T to determine the upper bound of number of anomalous machines in a safe scenario. That is to say, if ACOM on one anomalous machine can obtain information of more than T anomalous machines from WAN, it will alert user to potential high risk. If ACOM finds less than T anomalous machines from WAN, it concludes there is no cryptoworm attack and reports its judgement to user. An ant's goal is related to the threshold T. It is defined as the number of anomalous machines that the ant 
needs to collect during its travel. Let the value of goal be G, 
\begin{equation}
    G = T - P'.
\end{equation}
In equation (3), P' indicates the number of anomalous machines known to the local host that created the ant, and it is treated as the pheromone level. We will explain more details about pheromone in the next function ExchangeInformation(). The value of goal equals to the difference between threshold and pheromone because before a specific ant is created, some other ants may have travelled through this local host and deposit information about other anomalous machines observed during their traversals. As such, leveraging such information, this new ant will not need to start from scratch to reach the threshold. If the ant can find G anomalous machines from WAN, we think this machine is probably infected by cryptoworm. Otherwise, we report this machine is probably not infected. 
That is, ACOM will report our judgement according to ant's detection results.

The second thing the local host needs to tell the created ant is the home address. Home address is the IP address of this local host. With this address information, the ant could return and report detection results when it finishes its work.

The system parameter \textit{limit} stipulates that each ant can only travel through at most N machines. We set this limitation because we do not want the ant to go through so many machines that consumes a great amount of time and network resources.

\noindent
\textbf{Key Function 2: ExchangeInformation()}

As an ant arrives at a new machine, it exchanges information with the current local host so that both the ant and the current local host can enrich their knowledge about security condition in WAN. On one hand, ant tells local host a list contains all anomalous machines it has collected up to now as well as the count of anomalous machines which is considered as pheromone. This process is to mimic the behavior of ants in nature that leave pheromone trails on their way to food resources. On the other hand, local host tells ant its local detection result: whether it is anomalous or not. So, after exchanging information, ant may collect one more record while local host receives pheromone.

We also mimicked the property of pheromone that, it evaporates over time. We use this property because the machines do not need to keep very old information on them since the conditions of other machines in WAN may change over time. In our model, pheromone value remains unchanged in the first 10 seconds after it reaches the local host. Then, it decreases at a rate of 10\% per second. Suppose the original amount of pheromone is p, we can calculate pheromone p’ left on some machine after t seconds using this formula:

\begin{equation}
    p'(t) = \left \lfloor 0.9^{t-10} * p \right \rfloor , t \geq 10.
\end{equation}
Review the goal of each ant in function CreateAnt(), the value of p' we can achieve in equation (4) should be used as the variable p' in the equation (3) to calculate the goal of each ant when being created.

After exchanging information, the ant can decide whether it should go back home and report its detection result. If it has not finished its work, the local host should help ant decide direction, that is, which machine to go as the next stop.

\noindent
\textbf{Key Function 3: DecideDirection()}

In nature, ants decide their directions depending on the strength of pheromone trails ahead; In ACOM, the next destination of an ant is also decided depending on pheromone information left on the current local host. Since we want the ant to achieve its goal in shorter time if there exist some anomalous machines in WAN, the optimal direction of the ant should be an anomalous machine so that it can finish its work earlier.

To help an ant choose the next stop according to pheromone information on the current local host, our strategy is to assign weights to other machines that the current local host can contact with. Since the local host has pheromone information left by all passed ants, it has already known some anomalous machines in WAN. So, it should assign larger weights to these already known anomalous machines just like the already known shorter paths in nature having stronger pheromone trails. It assigns smaller weights to unknown machines just like uncertain paths to food sources in nature having weaker pheromone trails. In our implementation, the larger weights are set to 2 while the smaller weights are set to 1 to simply distinguish known anomalous machines and unknown machines. The stops which an ant has previously passed are assigned with weight 0 because the ant does not need to go back to the previous stops to gather information.

With weights set, current local host can calculate the possibility of each machine to be chosen as the next stop. The anomalous machines which have larger weights are more likely to be selected as destination of the ant. Suppose there are totally n machines in reach, the probability for some machine to be chosen is equal to the weight of this machine over the total weights of all machines in reach:

\begin{equation}
    probability(k) = \frac{weight(k)}{\sum_{i=1}^n weight(i)}, 1 \leq k \leq n.
\end{equation}

By this way, the next stop of the ant is decided in random but is not completely in random. The ant is more likely to be sent to an anomalous machine so that it can collect sufficient anomalous machines to prove a risky condition as soon as possible if there exist cryptoworm attack. Meanwhile, it is also possible that the ant can go to an undiscovered machine just like an ant in nature opening up a new path. Thus, we can guarantee that the information collected by ants are typical enough to conclude the current situation in network while very limited network resources and time will be consumed by ACOM.

\subsection{Broadcasting Mechanism}
While ACOM is designed for collecting security condition information from WAN, another network-level detection method called Broadcasting Mechanism (BM) is especially designed for detection in LAN. It exhaustively inquiries all machines in LAN and uses wisdom of the crowd to help user determine whether the local host is infected. This process does not consume too much network resource since the number of machines in LAN is limited, but it provides overall view of security condition in LAN.

Once BM is launched on a local host, it broadcasts the anomalous condition of the local host to all other machines in LAN meanwhile it receives this kind of information from other anomalous machines so that it can have a general idea about the number of anomalous machines in LAN at this point. Then, it generates a comprehensive report to tell user current security condition in LAN. For example, if there are totally 100 machines and 80 of them are anomalous, BM will generate a report saying that "80\% machines in LAN also experience anomalies, so your computer is in high risk of cryptoworm attack." Based on the reports from ACOM and BM, the user can make a judgement by himself(herself) about whether to treat his(her) computer as an infected machine.
\section{Evaluation of Network-Assisted Approaches}
In this section, we describe how we established a test environment in which 100 Docker containers are used to simulate a real-world network scenario and a Linux ransomware sample called GonnaCry \cite{GonnaCry} is applied on simulative infected machines to evaluate the performances of NAA. 

Although NAA is an integrated approach,
we compared the accuracy, message overheads and latency of local detection mechanism, ACOM and BM to verify whether network-level detection can improve local detection and to verify applicability of ACOM and BM in different scenarios. To distinguish the local detection mechanism used by ACOM and BM with the mechanism itself when treated as an independent mechanism, we name the independent local detection mechanism \textit{Direct Report} (DR). In the rest of this section, we will compare DR, ACOM and BM to have an comprehensive evaluation about the performance of each part of NAA. Note that, DR directly uses the detection result of local detection mechanism as the final result; ACOM is supported by the local detection mechanism and further uses the ACO algorithm to perform network-level detection to achieve a final report; BM also uses the local detection mechanism as a baseline and then collects information of all machines in simulative network to make a final report according to the number of anomalous machines.

\subsection{Experiment Environment}
Docker is a platform that provides resources and services for application development and test. It uses OS-level virtualization to deliver software in packages called containers. Containers can be considered as simplified virtual machines because each container has its own configuration files and libraries but is run by a single operating system kernel which results in fewer resources demands. Containers can communicate with each other through well-defined channels as well as maintaining isolated from one another. So, we use Docker containers to simulate the real-world network scenario instead of using virtual machines due to the functionality and simplification of containers. In our experiment, we established 100 containers, each of which is equipped with DR, ACOM and BM, to simulate a network environment containing 100 machines which can communicate with each other when it is needed. When testing a specific mechanism, we run this mechanism on all 100 containers for 10 times and observe its average performances.

To simulate the scenarios that some specified machines are attacked by ransomware, we run a Linux ransomware sample called GonnaCry on these specified containers and then execute a detection mechanism on each container to test its performances in this situation. GonnaCry employs a hybrid scheme which is utilized by most real-world ransomware nowadays combining asymmetric encryption and symmetric encryption together. To make the ransomware more secure from the attacker's perspective, GonnaCry contacts a remote server which keeps a pair of RSA keys for it, although the ransomware itself also has its own RSA key pair so that the victims cannot get the decryption key directly from their local hosts. The working procedure of GonnaCry is as following: The remote server generates a pair of RSA keys. The public key S\_pub is hardcoded in GonnaCry while the private key S\_priv is preserved on the remote server. When GonnaCry starts to work, it generates its own RSA key pair on the local host. The public key is called C\_pub and the private key is called C\_priv. Then, it uses AES cipher to encrypt the local private key C\_priv with the server’s public key S\_pub and also uses AES cipher to encrypt target files with local public key C\_pub. In this case, if someone wants to recover these encrypted files, he/she needs to get the server’s private key S\_priv first to recover the local private key C\_priv so that he/she can use C\_priv to decrypt files. Since the server’s private key S\_priv is stored on the remote server, the victim has to pay the ransom to obtain this key. We apply GonnaCry on simulative infected machines due to its realism.

We respectively simulated 11 different scenarios with increasing numbers of infected machines and decreasing numbers of safe machines while the total number is always 100. In each scenario, we respectively apply three different mechanisms on containers and test 10 times to achieve reasonable average results of accuracy, message overhead and latency.

To determine the value of limit N and threshold T, we tried many different values under this 100-machine scenario. Finally, we decided that N = 20 and T = 3 because this setting contributes a best balance between accuracy and efficiency which considers both time consumption and network resource consumption.

\subsection{Accuracy}
Accuracy is defined as the correctly reported cases out of overall cases, that is, accuracy = (true positives + true negatives) / (true positives + false positives + true negatives + false negatives). Since BM just reports the fact it observed using wisdom of the crowd instead of reporting its own judgement, we only evaluate the accuracy of DR and ACOM. The result is shown in Figure 6, the x-axis represents the number of infected machines, the y-axis indicates accuracy of DR and ACOM.

\begin{figure}[!htb]
	\centering
		\includegraphics[width=3.3in]{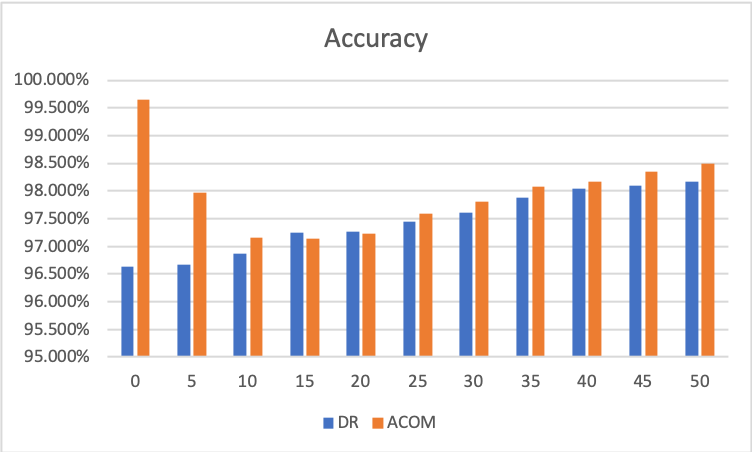}
	\caption{Accuracy Comparison of ACOM and DR}
\end{figure}

We can observe that ACOM has greater advantage over DR when there are only a few infected machines. As the number of infected machines increases, although ACOM does not have evident superiority, it is still more accurate than DR in most cases. This test result proves that the network-level detection can help improve accuracy of local detection. Plus the comprehensive report from BM, user can make an even more precise decision about whether the local host is attacked by ransomware. If the ransomware is a cryptoworm, it can be detected at very beginning if NAA is deployed due to high accuracy of ACOM at the time that only a few machines are infected.

\subsection{Message Overheads}
Message overhead is another important factor in consideration since we do not want to cause too much network resource consumption during the process of ransomware detection. If a ransomware detection approach produces huge resource consumption which is heavier than the damage of ransomware itself, it should not be put into practice. It is obvious that these three mechanisms we put forward will not cause huge resource consumption compared with the expensive extortion fee of ransomware, but we still want to figure out their message overhead to see which mechanism is optimal from this aspect. We define message overhead as the extra messages produced by ransomware detection approaches. In ACOM, machines need to send and receive ants during the detection process. In BM, machines need to send and receive news about whether a specific machine is anomalous or not. So, both ACOM and BM produce extra messages when they are running. Figure 7 shows the message overhead of Dr, BM and ACOM. The x-axis indicates the number of infected machines and the y-axis indicates the number of messages being produced during each detection process.

\begin{figure}[!htb]
	\centering
		\includegraphics[width=3.3in]{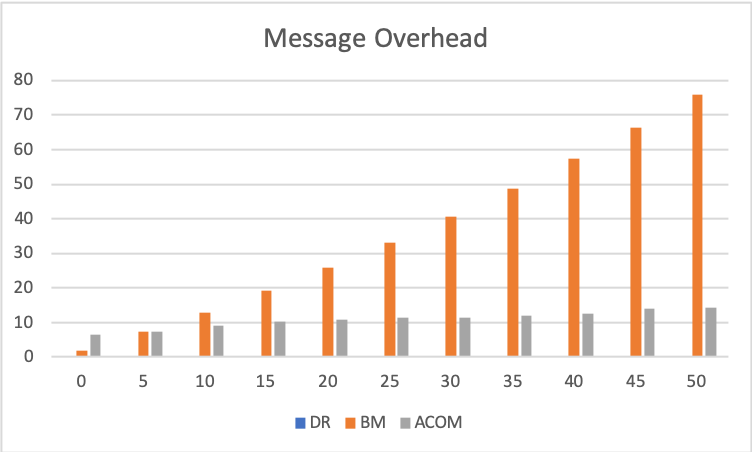}
	\caption{Message overhead of three mechanisms.}
\end{figure}

We can observe that DR performs best when coming across message overhead measurement because it directly uses the detection results of local detection mechanism which does not produce any additional messages. BM produces more message overheads than ACOM does in most cases. As the number of infected machines increases, the message overhead of BM drastically grows while that of ACOM slightly grows. The reason is that, BM requires each anomalous machine to send messages to all peers while ACOM only allows each ant to go through at most 20 machines. Thus, apply BM to LAN scenario is a reasonable arrangement from message overhead's perspective since there are limited machines in LAN making the message overhead of BM countable.

\subsection{Latency}
Latency is the time duration that each mechanism needs to complete its task. We calculated the average latency on all machines  in each test. Figure 8 shows the latency of DR, BM and ACOM. The x-axis indicates the number of infected machines and the y-axis indicates the average seconds that each mechanism consumes during its work.

\begin{figure}[!htb]
	\centering
		\includegraphics[width=3.3in]{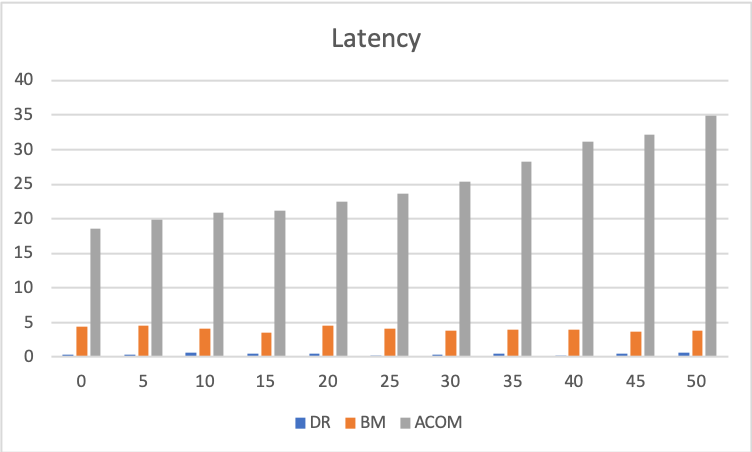}
	\caption{Latency of three mechanisms.}
\end{figure}

As for DR, its latency is approximately 0 because it only does local detection which can be completed in very short time. As for BM, no matter how many victims exist, the anomalous machines always broadcast a message and receive messages from other anomalous peers and then a report is sent to user depending on the number of anomalous machines in LAN. All machines work in parallel following the above procedure, which makes the runtime of all machines be similar to the runtime of one randomly picked machine. So, the average latency of BM only has a little fluctuation as the number of infected machines increases. As for ACOM, each anomalous machine creates an ant that goes through at least 3 machines one after one. As the number of infected machines increases, more ants will be created which makes average runtime increase. So, ACOM has the worst latency among three integrated approaches while Direct Report almost has no latency. However, the high latency of ACOM does not do extra damage to infected machines because all suspicious tasks are suspended before ACOM is launched so that ransomware cannot encrypt files when network-level detection is working. 

\subsection{Loss Assessment}
In this section, we estimated the damages that a ransomware can cause on a machine before it is detected by our ransomware detection approach NAA. That is, how many files can be encrypted before the ransomware is detected and terminated.

We can learn from the test results shown above that ACOM has relatively long delay before reporting our diagnosis to user. However, it does not result in additional damage because before ACOM is launched, all suspicious tasks are suspended until user takes further actions. So, the number of files being encrypted during the process of local detection is exactly the losses of this machine. Figure 9 shows the average number of encrypted files on a victim machine if NAA is applied on. The x-axis is the number of infected machines in LAN, the y-axis is the average number of encrypted files.

\begin{figure}[!htb]
	\centering
	\includegraphics[width=3.3in]{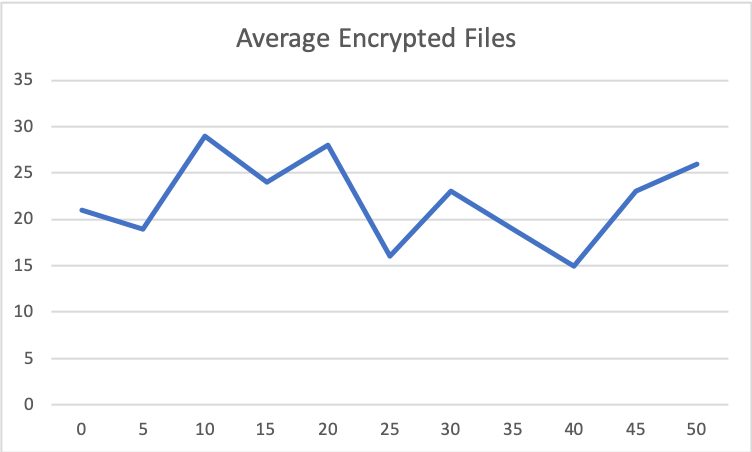}
	\caption{Average number of encrypted files.}
\end{figure}

We can observe that no matter how many machines are infected, the number of encrypted files on each machine ranges from 15 to 30, which is acceptable loss owe to the quick job of our local detection mechanism. \\

Based on our evaluation results concerning accuracy, message overheads, latency and loss assessment, we find that network-level detection can indeed help improve the accuracy of local detection. From message overhead's point of view, ACOM is applicable to WAN scenario while BM is applicable to LAN scenario for network-level detection. Moreover, NAA provides good performance especially for detecting cryptoworm attack since our network-level detection can provide user with very accurate alert in the early stage of cryptoworm attack.
\section{Conclusion and Future Work}
In this paper, we propose a network-assisted approach called NAA for ransomware detection which combines local detection and network-level detection together. We first describe a local detection mechanism which  uses three local features to judgement whether the local host is anomalous. In network-level detection, we implement ACOM to efficiently collect information in WAN scenario and put forward BM which exhaustively inquires all machines in LAN. Then, the network-level detection uses wisdom of the crowd to provide user with a comprehensive report so that user can easily make his(her) judgement based on the information we offered. To evaluate our approach, we use docker to establish the experiment environment and use GonnaCry to simulate ransomware attack. The test results show that NAA is more accurate than local only detection and is especially applicable for cryptoworm detection meanwhile the loss of files during the working procedure of NAA is acceptable.

However, due to the limited resource of Linux ransomware sample, we only used GonnaCry to simulate ransomware attack in our evaluation experiments. In the future, we will test the performance of NAA using some other Linux ransomware samples especially Linux cryptoworm samples when they are accessible.

\section{Acknowledgement}

This research was developed with funding from the Defense Advanced Research Projects Agency (DARPA). The views, opinions and/or findings expressed are those of the author and should not be interpreted as representing the official views or policies of the Department of Defense or the U.S. Government.

Distribution Statement ``A" (Approved for Public Release, Distribution Unlimited). 

\bibliographystyle{plain}
\bibliography{paper}
\end{document}